# Genomic Sequence Diversity and Population Structure of *Saccharomyces cerevisiae* Assessed by RAD-seq -

Gareth A. Cromie [1*], Katie E. Hyma[2*], Catherine L. Ludlow[1], Cecilia Garmendia-Torres[1], Teresa L Gilbert[‡], Patrick May[1,3], Angela A. Huang[4], Aimée M. Dudley[1,5], Justin C. Fay[6]

[1] Institute for Systems Biology, Seattle, WA, USA

[2] Bioinformatics Facility (CBSU), Institute for Biotechnology, Cornell University, Ithaca, NY, USA

[3] Luxembourg Centre for Systems Biomedicine, University of Luxembourg, Esch-sur-Alzette, Luxembourg

[4] University of Pennsylvania, PA, USA

[5] Corresponding author

[6] Department of Genetics, Washington University, St. Louis, MO, USA

* These authors contributed equally to this work.

[‡] Deceased

**Correspondence:**

Aimée M. Dudley (aimee.dudley@systemsbiology.org)

Institute for Systems Biology

401 Terry Avenue North

Seattle, WA 98109

Tel: (206) 732-1214

Fax: (206) 732-1299

## Abstract

The budding yeast *Saccharomyces cerevisiae* is important for human food production and as a model organism for biological research. The genetic diversity contained in the global population of yeast strains represents a valuable resource for a number of fields, including genetics, bioengineering, and studies of evolution and population structure. Here, we apply a multiplexed, reduced genome sequencing strategy (known as RAD-seq) to genotype a large collection of *S. cerevisiae* strains, isolated from a wide range of geographical locations and environmental niches. The method permits the sequencing of the same 1% of all genomes, producing a multiple sequence alignment of 116,880 bases across 262 strains. We find diversity among these strains is principally organized by geography, with European, North American, Asian and African/S. E. Asian populations defining the major axes of genetic variation. At a finer scale, small groups of strains from cacao, olives and sake are defined by unique variants not present in other strains. One population, containing strains from a variety of fermentations, exhibits high levels of heterozygosity and mixtures of alleles from European and Asian populations, indicating an admixed origin for this group. In the context of this global diversity, we demonstrate that a collection of seven strains commonly used in the laboratory encompasses only one quarter of the genetic diversity present in the full collection of strains, underscoring the relatively limited genetic diversity captured by the current set of lab strains. We propose a model of geographic differentiation followed by human-associated admixture, primarily between European and Asian populations and more recently between European and North American populations. The large collection of genotyped yeast strains characterized here will provide a useful resource for the broad community of yeast researchers.

## Author Summary

Baker's yeast, *Saccharomyces cerevisiae*, is a simple single-celled fungus important for producing fermented foodstuffs such as bread, wine and beer. *S. cerevisiae* is also widely used in the laboratory where it has served as a powerful model for understanding basic cellular processes that are common to many organisms, including humans. However, compared to the relatively small set of laboratory strains, the global population of yeast represents a much more extensive source of DNA sequence variation. To explore the genetic diversity of the global population of yeast, we used a cost effective genotyping method that permits the sequencing of a common ~1% of every strain's genome. Our results suggest that major subpopulations of yeast correspond to geography (Asia, Southeast Asia and Africa, Europe, North America) and confirm that human activity has dispersed strains around the globe. Strains used in the laboratory are mostly related to European strains and represent only about one quarter of the sequence diversity in the whole population.

## Introduction

The budding yeast *Saccharomyces cerevisiae* has been used by humans for thousands of years to produce food and drink products that rely on fermentation, such as bread, beer, sake and wine. In recent decades, *S. cerevisiae* has also proved to be a powerful model organism for the study of eukaryotic biology. Many cellular processes are highly conserved from yeast to higher eukaryotes, and the ease of propagating and manipulating this simple single-celled organism has led to its use in a wide variety of research areas. Yeast is a particularly powerful model system for genetics. *S. cerevisiae* was the first eukaryote to have its genome fully sequenced [1], and nearly 85% of yeast genes have functions that are characterized to some extent, a much higher fraction than any other eukaryotic organism. *S. cerevisiae* has also been used to develop many of the modern high-throughput tools to study the genome, transcriptome and proteome [2]. Together, these advantages have positioned *S. cerevisiae* as the eukaryotic organism that we have come closest to understanding at a global or systems level.

Research using *S. cerevisiae* has traditionally focused on a small number of well-studied laboratory strains. However, in recent years, interest in natural isolates has increased as it has become clear that many non-laboratory strains (including those adapted to various food/industrial processes) have properties absent from the lab strains, such as the ability of several wine strains to ferment xylose [3]. The wider population of yeast strains represents a deep pool of naturally occurring sequence variation that has been leveraged to investigate the genetic architecture of polygenic traits [4-10]. In addition, the polymorphisms that are observed in the global yeast population have been acted upon by evolution, making this set of sequences a powerful tool for investigating protein and regulatory sequence function as well as evolution [11].

Understanding the genetic diversity of yeast is therefore relevant to both the food/industrial roles of yeast and its role as a model organism in scientific research.

The question of the global population structure of *S. cerevisiae* is itself an ongoing topic of research. Several publications in the past few years have explored the genetic diversity and population structure of yeast using techniques such as multi-gene sequencing [12-15], whole genome sequencing [16], tiling array hybridization [17] and microsatellite comparisons [18-21]. These studies demonstrated that *S. cerevisiae* is not a purely domesticated organism but can be isolated from a variety of natural environments around the globe. While there appears to be some clustering of yeast genotypes by geography [16], it also appears that yeast involved in particular human food-industrial processes are often genetically similar to one another. For example, wine strains isolated from around the world display a very high degree of sequence similarity [12, 16, 17, 21]. Unfortunately, several of the most diverged groups identified in these studies were represented by relatively small numbers of strains, suggesting that analysis of additional strains might help clarify the structure of global yeast diversity.

Whole genome sequencing of a large, diverse set of individuals is the most comprehensive approach to exploring the population structure and genetic diversity of an organism. However, despite the falling costs of DNA sequencing, complete genome sequencing of several hundred yeast strains is still a significant expense. In contrast, methods that compare strains by genotyping relatively small numbers of loci, such as microsatellites or a small number of genes [12], are less expensive, but the results may not reflect the relationships between strains genome-wide. A genome reduction strategy referred to as restriction site associated sequencing (RAD-seq) [22, 23] directs sequence reads to genomic locations adjacent to particular restriction sites. However, because most restriction sites are common across strains of the same species, nearly the same subset of every genome is sequenced. Finally, because the sequenced regions are spread across the genome, comparisons should reflect the pattern that would be observed from comparisons of whole genome sequences. Thus, the RAD-seq approach should accurately capture the relationships seen with whole genome sequencing, but at a fraction of the cost.

In this work, we apply a multiplexed RAD-seq reduced genome sequencing strategy to explore genetic diversity and population structure in *S. cerevisiae*. Using this approach we sequenced more than 200 strains over ~1% of the yeast genome. The strains include multiple representatives from six continents, 38 different countries and were isolated from disparate sources, including fruits, insects, plants, soil and a variety of human fermentations, such as ragi, togwa, cacao, and olives. From analysis of the resulting multiple alignment, we observed a clear geographical stratification of strains along with evidence of admixture between populations and human-associated strain dispersal.

## Results

In an effort to expand the number and diversity of characterized *S. cerevisiae* strains available to the yeast community, we assembled and characterized a collection of >200 strains (Materials and Methods, Table S1). This strain set covers a diverse range of ecological niches and geographical locations, including strains used in previous studies of yeast global and local population structure [12, 16-19] and strains with published whole genome sequence (WGS) data. We sequenced each of these strains using a RAD-seq strategy to produce an initial multiple alignment (Materials and Methods). Strains with published WGS data were then added to the alignment to facilitate comparison between the results generated using WGS and RAD-seq data (Materials and Methods). The final dataset contained 262 strains genotyped across 116,880 base positions of which 5,868 sites were polymorphic (Supplemental Dataset 1).

### Genetic Relationship Among Strains

To visualize the phylogenetic relationships between the strains, we generated a neighbor-joining tree from the reduced genome multiple alignment (Figure 1 and Supplemental Dataset 2). The tree agrees well with the geographic origins of the strains and, for the subset of strains in common, is also consistent with a previous study that used WGS [16]. The most distantly related populations of North America, Europe, South Asia and East Asia form clear and well-separated clusters on our tree. We also identified a small isolated cluster of strains from Ghana involved in cacao fermentation and another discrete cluster of strains from the Philippines.

A clear exception to this geographical stratification is the dispersal of European/wine strains around the globe, a result that is also consistent with the previous study [16]. We identified two clusters of strains that appear closely related to the European/wine cluster, one isolated from European olives and another consisting primarily of a collection of environmental isolates from New Zealand [19]. Results for this second group are consistent with the hypothesis that the strains largely reflect a population brought to New Zealand as a consequence of European settlement. Together with the main “European/Wine” cluster, these two groups of strains appear to identify a “greater-European” region of the tree.

Strains isolated from North America fell into two highly diverged regions of the tree. One set of strains (Figure 1, “North America Wild”) defines a cluster of strains almost universally isolated from North America (largely environmental samples from soil and vegetation). The second set is genetically similar to the European/wine strains, with strains scattered within the main European/wine cluster and related groups (Figure 1 and Table S1). There are also a small number of strains isolated from North American environments in the “New Zealand” cluster. As previously observed [24], North American strains isolated from even the same locale (e.g. a single vineyard) split into subsets from both the North American Wild cluster and greater-European regions of the tree. These results are consistent with the assertion that in many locations across North America (particularly vineyards), a native population of yeast strains coexists sympatrically with a population introduced by European settlement [24].

Another instance in which highly diverged strains were isolated from a single small geographical location is provided by the set of strains isolated from “Evolution Canyon”, a well-studied location in Mount Carmel National Park of Israel [18]. These strains fell into one large and two smaller clusters on the tree (Figure1, Israel 1, Israel 2, and a third cluster within a diverse set of strains labeled “Mixed”). The genomic diversity of these strains is remarkable, given that they were collected within a few hundred meters of each other.

## Strains Widely Used in the Laboratory

Included in the multiple alignment and phylogenetic tree is a group of seven strains widely used in the laboratory (S288c, W303, RM11-1a, FL100, Sigma 1278b, SK1, Y55), several of which are known to be closely related [25]. The strains SK1 and Y55 are closely related to the West African cluster while S288c, FL100 and W303 are related and close to the European/Wine cluster. The position of these strains on the tree agrees with two previous studies [16, 17], both of which described the limited sequence diversity of the lab strains. For example, none of the commonly used lab strains are derived from certain major populations, including the Asian group and the North America Wild group (Figure 1). Together, these results suggest that the total sequence diversity of the yeast global population is poorly sampled by this set of strains in common laboratory use.

To compare the total sequence diversity captured by the full set of 262 strains relative to that present in the subset of laboratory strains, we analyzed all alleles (defined as single base pair polymorphisms) that occurred in more than one strain. Alleles found in only one strain (singletons) were ignored to reduce the effect of sequencing errors, as were heterozygous calls. The results show a total of 3,321 polymorphic loci with 6,680 total alleles (3,283 bi-allelic, 38 tri-allelic, and 0 tetra-allelic positions). Only 1,703 of these 6,680 alleles were observed in the set of lab strains, and thus the set of strains assembled in our panel represents a significant increase (~4-fold) in sequence diversity over the set of lab strains.

## Population Structure

The infrequent sexual cycle of *S. cerevisiae*, combined with its high rate of self-mating, promotes the establishment of strong population structure and enables clonal expansion of admixed populations. To infer population structure and admixture between populations while accounting for selfing, we applied a Monte Carlo Markov chain algorithm, InStruct [26], to the 759 sites with an allele frequency of 10% or more. On the basis of the deviance information criterion, we inferred the most likely number of populations to be nine (Materials and Methods) and labeled each population by the most common geographic location and/or substrate from which the strains were originally isolated (Table S2). The relevant genotypes of each strain along with their inferred population ancestry are shown in Figure 2 and Table S1. The nine populations consist of two North American oak populations, an Asian food and drink population, a European wine & olive population, an African/S. E. Asian population, a New Zealand population, an Israeli population and two populations associated with industrial/food processes. These populations match well with the major groupings seen on the phylogenetic tree, with the two North American populations identified by InStruct corresponding to the "North America Wild" grouping (Figure 1 and Figure S1). It is notable that these two subdivisions do not reflect a clear geographic pattern within North America (Figure 2 and Table S1). The New Zealand population clearly shares many alleles with the European strains, but harbors a small number of sites that make it unique. One of the two human-associated groups contains the majority of laboratory strains, emphasizing the uneven sampling of yeast populations represented by the set of laboratory strains.

## Admixture

For each population, strains were observed with high levels of ancestry to that population. However, 38% of strains showed appreciable levels of admixture, defined as less than 80% ancestry from a single population. To assess the overall coincidence of mixture between pairs of populations we tabulated the number of strains with at least 20% ancestry from each pair of populations (Figure 3). Most admixed strains involved the European, Asian or African populations. However, not all pairs of populations were equally likely to admix. Admixture was detected between the European population and the first North American (InStruct #1), but not the second North American (InStruct #2) populations. More generally, admixture with the two North American populations was largely restricted to the African and European populations or to admixture between the two populations themselves. Like the European population, the Asian population showed admixture with most other groups. The two human-associated populations were largely admixed with either the Asian or European populations. Finally, the New Zealand population only admixed with the European population, and the Israeli population was largely admixed with the Asian and one of the human-associated populations.

## Heterozygosity

Matings within or between populations can result in strains with a large proportion of heterozygous sites. Most strains in this study had zero or a relatively small number of such sites. These strains could be naturally occurring homozygotes, haploids, or converted to homozygous diploids, a standard practice in some laboratories. However, we did identify 65 strains with more than 20 heterozygous sites (Table S1). The two strains with the highest number of heterozygous sites DCM6 (n= 305) and DCM21 (n= 288) were isolated from cherry trees in North America and appear to be hybrids between the European and North American populations [24]. Other strains with a large number of heterozygous sites (Table S1) were also isolated from fruit-related sources, including three from cacao fermentations, one from banana fruit, one from fruit juice and one from a spontaneous grape juice fermentation. Across these 65 strains, 42 also exhibit notable admixture, defined by less than 80% ancestry from a single population. The proportion of heterozygous strains exhibiting appreciable admixture (65%) is significantly greater (Fisher's Exact Test, $P = 1.5\text{x}10^{-4}$) than strains with little or no heterozygosity (38%), suggesting that heterozygosity was derived in part by admixture between populations. Among the heterozygous strains, the highest proportion of ancestry comes from one of the human-associated populations (#4, 31%), followed by the European (20%), Asian (17%) and African (14%) populations. To examine rates of heterozygosity across populations, we compared expected to observed heterozygosity within each population (Table S1). While most populations exhibit a deficit of observed, compared to expected heterozygosity, the two human-associated populations show noticeably more heterozygosity than the other populations.

## Relatedness Between Populations

Whereas heterozygosity and admixture can provide information about strain ancestry, relatedness between populations can provide information about the history of entire populations, some of which may themselves be derived from historical admixture events. To examine relatedness between populations, we applied multidimensional scaling (Materials and Methods) to the entire dataset (Figure 4). The first principal coordinate differentiates the European population from the other populations; the second principal coordinate distinguishes the two North American populations from the Asian population; and the third principal coordinate differentiates the African/S. E. Asian population from the others. The remaining populations and most of the admixed strains lie between these four major groups (Figure 4). Consistent with their positions on the neighbor-joining tree (Figure 1) and their genotypes (Figure 2), the New Zealand and Israeli populations are most closely related to the European population, and the two human-associated populations lie between the European and Asian populations. The results, combined with its high rates of heterozygosity, also suggest that the first human-associated population (population #4) appears to be a recently derived population originating from hybrids between the European and Asian populations.

### Subpopulations

Low frequency alleles (<10%) can sometimes define subpopulations not captured by inference of population structure based on common alleles. Two-dimensional hierarchical clustering of the low frequency alleles identified a number of such subgroups (Figure 5 and Table S1). These subgroups include a previously described Malaysian/Bertram Palm population [16], but also groups of strains from Philippines/Nipa palm, togwa, olives, sake and cacao. While the number of strains in each group is small, the number of sites defining the groups is not. In support of these groups representing populations that have been isolated from other populations for some time, many of the rare variants that define these groups are not present in other strains but, in at least some cases, are variable within the subgroup. Interestingly, the subpopulations defined by the largest numbers of alleles are strains with primary membership to the African/ S. E. Asian population, suggesting that there may be undiscovered subpopulations and diversity among strains of African or Southeast Asian origin.

## Discussion

Genetic variation within *S. cerevisiae* has been shaped by a complex history, influenced by human-associated dispersal and admixture. Understanding this history and the resulting patterns of diversity is important for capturing and harnessing its fermentative capabilities as well as for quantitative and population genetics research. In this study, we used a reduced genome sequencing strategy to characterize the genetic diversity among a global sample of 262 strains isolated from a wide range of ecological habitats and environmental substrates. Our findings indicate that the major axes of differentiation correspond to broad geographic regions. In addition to previously described populations and patterns of differentiation [12, 16, 17, 19, 21, 27], two new patterns indicative of human influence also emerge. First, we find a population represented by multiple human-associated strains that contains a mixture of European and Asian alleles. Second, we find human-associated subpopulations from togwa, olive, cacao and sake fermentations that are defined by a unique set of variants not present elsewhere in the global population. While inferences of population structure can depend on sampling, indeed our analysis points to areas of uncertainty, the structure of *S. cerevisiae* described here is based on the largest collection of strains typed across the genome. This work also provides a foundation for studying the genetic underpinnings of complex traits, the origin and evolution of strains used by humans, and the relationships between such traits and population history.

## Geographic Differentiation

A major unanswered question in the study of yeast population structure has been the relative importance of geography versus ecological niche. While the strains in this study were isolated from many different ecological habitats, a number of lines of evidence suggest that the groups they form are defined better by geographic differentiation than by ecological niche. The two North American populations contain predominantly oak-associated strains, but they also contain strains from plants and insects. Similarly, the European population contains primarily vineyard-associated strains, but also contains a number of European soil and clinical isolates [16]. The Asian population also includes strains isolated from multiple countries and several different habitats, including strains used in Sake fermentation and several strains isolated from food. The Asian population shares many alleles with the North American populations, but is genetically distinct and includes only a handful of strains from outside of Asia. What is less clear is how this Asian population is related to a number of diverged lineages represented by strains from primeval and secondary forests in China [13].

In comparison to the European, North American and Asian populations, the African/S. E. Asian population is not as well defined. Most of the strains are inferred to have mixed ancestry, and the strains that are most representative of the population (>80% ancestry) combine previously separated populations [16] of West Africa and Malaysia, two populations that are also separated on our tree (Figure 1). Because the trees are consistent, the different results of the two population analyses could be a result of differences in the methods of analysis (e.g. Structure versus InStruct), the larger number of strains used in this study, or the larger number of sites used by Liti et al. [16].

## Admixture

Evidence of admixture was seen in a large fraction of strains and in every population. While admixture is most common among the European, African and Asian populations (Figure 3), the smaller number of admixed strains from the North America and New Zealand populations may represent the more recent establishment of European strains in these locations or may be related to the frequency of mating in the oak tree or soil environment. Some of the admixed strains also exhibit high rates of heterozygosity, indicating a relatively recent mating between strains with different ancestries. Interestingly, many of the heterozygous strains were isolated from fruits or orchards, an observation that is consistent with the isolation of admixed (mosaic) strains from fruits and orchards in China [13].

Because yeast can grow asexually, entire populations can arise as a consequence of even rare admixture events. The two human-associated populations bear a strong signature of an admixed origin as they carry alleles from both European and Asian populations and lie between these two groups in the principal coordinate analysis (Figure 2 and Figure 4). Human-associated population #4 bears the additional signature of high rates of heterozygosity, implying relatively recent mating events in the origin of this group. In contrast, human-associated population #5 harbors fewer heterozygous strains, but also contains multiple laboratory strains (Sigma 1278b, FL100, W303, S288C and FY4), some of which show mosaic patterns across their genome indicative of an admixed origin [16, 25, 28].

The New Zealand and Israeli populations may also have an admixed origin. These two populations carry a large subset of the European alleles, similar to many of the admixed European strains, but also carry a small number of alleles present at high frequency in the North American or Asian populations. This pattern is consistent with New Zealand and Israeli populations being derived from an admixture event between the European and these other populations followed by clonal (or nearly clonal) expansion. However, the New Zealand and Israeli populations also carry a small number of alleles that are not present in either the North American or Asian populations (Figure 2). This raises the possibility that the New Zealand and Israeli populations were derived from admixture between the European and as yet undiscovered populations, or instead, rather than derived from an admixture event, that they represent lineages with roots in an ancestral European population (similar to the "Olive" grouping). The diversity of strains sampled from Evolution Canyon in Israeli is particularly notable. Of the 15 Israeli strains, seven define the nearly clonal Israeli population, three are assigned with 100% ancestry to the human-associated population #4, and four show comparable percentages of ancestry from the Asian, Israeli and human-associated (#5) populations.

## Derived Subpopulations

The use of common sites to infer population structure eliminates the detection of small populations defined by rare variants. With clustering based solely on rare variants, we identified a number of such subpopulations (Figure 5). Although many of these groups were isolated from human-associated fermentations, the number of strains is too small to clearly indicate whether they are related by geographic or environmental origin. For example, the olive strain group contains isolates from Spanish olives imported to Seattle and one from olives in Spain. Yet, this group does not contain two strains isolated from the brine of olives from Mexico and one from an olive tree in California. The two North American groups contain strains from different states, and the togwa and cacao strains were each sampled from the same country. While some of these subpopulations may be the result of recently expanded clones, several of them are defined by sites that are variable within the subpopulation. This latter observation points to the establishment of small groups that have remained isolated due to either geographic or ecological barriers to gene flow.

## Prospects for Future Studies

As our understanding of *S. cerevisiae* population history increases, so does the need to incorporate such information into quantitative and population genetic studies. Our results highlight the complex relationships between strains and populations, but also characterize a set of strains and sequences that can be used by the community. Using whole-genome sequencing or a reduced genome sequencing strategy, such as the RAD-seq method used here, new strains can be readily placed in the context of global population structure. We anticipate that new genetic diversity will be discovered, particularly in Africa for which we found less certain relationships and a number of derived subpopulations. Our results may also prove useful to studies of existing strains, either by controlling for population history in genome-wide association studies or by aiding the selection of strains for linkage analysis. In both cases, strain choice is an important consideration as the results can depend on what variation is captured and the structure of this variation across strains. While many quantitative genetic studies have been based on crosses with laboratory strains, our results underscore the presence of additional variation that is available beyond those strains. Finally, the global diversity and increased variation uncovered by our study highlight the potential for identifying novel properties which could prove valuable to the improvement of existing strains or the engineering of new strains for use in industrial fermentations.

## Materials and Methods

The *S. cerevisiae* strains analyzed in this study were obtained from a variety of sources, including the Phaff Yeast Culture Collection (http://www.phaffcollection.org/), the ARS (NRRL) Culture Collection (http://nrrl.ncaur.usda.gov/), published strains from individual laboratories or our own isolates from wild or domesticated sources. Details, including references and information about strain requests, are included in Table S1. While analyzing the data we came across a small number of anomalies, such as two dissimilar genome sequences for strain 322134S. These are likely to represent errors in strain labeling.

### Yeast Isolation

Soil, bark and leaves or food samples were bathed in medium consisting of 2 g/L Yeast Nitrogen Base without amino acids (Difco, BD,), 5 g/L ammonium sulfate and 80 g/L glucose. Chloramphenicol (30 µg/ml) and carbenicillin (50 µg/ml) were added to the medium to suppress bacterial growth and cultures were incubated at 30ºC. When necessary to suppress mold overgrowth, cultures were sub-cultured to liquid medium containing 1-5% ethanol. Cultures were examined by microscopy at 3 days and 10 days, and those harboring budding yeast were plated onto CHROMagar Candida (DRG International, Inc.) and incubated at 30ºC for 3-5 days. CHROMagar Candida is a culture medium containing proprietary chromogenic substrates that can aid the identification of clinically important yeast [29]. On CHROMagar Candida, *S. cerevisiae* colonies are known to range in hue from white to lavender to deep purple with most exhibiting the "purple" phenotype (Ludlow and Dudley, unpublished result; [30]). Colonies exhibiting these color phenotypes were picked and saved for further study.

## RAD-Sequencing and Alignment

A subset of strains were RAD-sequenced previously [24]. For the rest, RAD-sequencing was carried out as previously described [24, 31]. Briefly, yeast genomic DNA was extracted in 96-well format and fragmented by restriction enzyme digestion with *Mfe*I and *Mbo*I. P1 and P2 Adaptors were then ligated onto the fragments. The P1 adaptor contains the Illumina PCR Forward sequencing primer sequence followed by one of 48 unique 4- nucleotide barcodes and finally the *Mfe*I overhang sequence. The P2 adaptor contains the Illumina PCR Reverse primer sequence followed by the *Mbo*I overhang sequence. After ligation, the barcoded ligation products were pooled, concentrated, and size selected on agarose gels, with fragments from 150 to 500 base pairs extracted from the gel. Gel-extracted DNA was further pooled to multiplex 48 uniquely barcoded samples in one sequencing library. The multiplexed DNA library was then enriched with a PCR reaction using Illumina PCR Forward and Reverse primers. Sequencing runs were performed on the Genome Analyzer IIx (Illumina) for 40 base pair single-end reads, with one library of 48 multiplexed samples per flow cell lane, yielding 20-40 million reads. The read sequences generated for this study are available at <Upon acceptance, sequences will be deposited to DRYAD and SRA>, and for the subset of strains that were RAD-sequenced previously, DRYAD entry doi:10.5061/dryad.g5jj6.

Multiple sequence alignments were generated by mapping reads to the S288c reference genome (chromosome accessions: NC_001133.8, NC_001134.7, NC_001135.4, NC_001136.8, NC_001137.2, NC_001138.4, NC_001139.8, NC_001140.5, NC_001141.1, NC_001142.7, NC_001143.7, NC_001144.4, NC_001133.8, NC_001133.8, NC_001133.8, NC_001133.8) and generating consensus reduced-genome sequences for each strain. The tagged reads were split into strain-pools by their 4 base prefix barcodes. Reads with N's or with Phred quality scores less than 20 in the barcode sequence were removed. Any reads with more than 2 Ns outside the barcode were also removed. Reads were aligned to the S288c reference using BWA (V0.5.8, [32]) with 6 or fewer mismatches tolerated. Samtools (V0.1.8, [33]) was then used to generate a pileup from the aligned reads using the “pileup” command and the “-c” parameter . Base calls were retained if they had a consensus quality greater than 20. Positions with root mean squared (RMS) mapping qualities less than 15 and insertion/deletion polymorphisms were ignored. After filtering there was an average of 209,765 bp for each strain. Sequences from each strain were combined into a multiple sequence alignment via their common alignment to the S288c genome. Sites with more than 10% missing data were removed, resulting in a multiple sequence alignment of 116,880 base pairs.

## Whole Genome Sequencing Alignment

Previously-generated whole genome sequences (WGS) were incorporated into the RAD-seq dataset for population genetic analysis. For genomes with an S288c NCBI coordinate system, sequences were extracted directly based on S288c reference coordinates. For genomes using an alternative coordinate system (SGRP), blat was used to convert from the S288c NCBI reference coordinates to the alternative coordinate system prior to extracting sequences. For assembled genomes without S288c alignments, coordinates were obtained by blast. A fasta file of the S288c reference sequence was generated for each contiguous segment in the multiple sequence alignment. The resulting files were used to query each genome assembly using blast. When quality scores were available, sites with sequence quality less than 20 were converted to “N,” prior to blasting or following sequence retrieval.

## Duplicated Strains

Some strains were sequenced by both WGS and RAD-seq. For duplicate strains with pairwise divergence less than 0.0005 substitutions per site, excluding singleton alleles (i.e. found in only 1 strain), only the RAD-seq data was retained for analysis. For duplicate strains that exceeded the threshold, both RAD-seq and WGS data were retained and strain names were labeled with an "r" and "g", respectively.

### Population Analysis

Neighbor-joining phylogenetic tree construction was carried out using MEGA [34] (V5.0), based on P-distance with pairwise deletion. Population structure was inferred using InStruct [26]. Because InStruct failed to converge using all sites, it was instead run on 759 sites with allele frequency greater than or equal to 10%. Polymorphic sites were made biallelic by treating third alleles as missing data. InStruct was run with the parameters "-u 40000 -b 20000 -t 10 -c 10 -sl 0.95 -a 0 -g 1 -r 1000 -p 2 -v 2" with K (number of populations) ranging from 3 to 15. While the lowest deviance information criterion (DIC) was obtained from a chain with K = 15, there was substantial variation among independent chains. We chose K = 9 as the optimal model to work with based on the average DIC for K = 10 being nearly identical to that of K = 9 and subsequent drops in DIC for larger values of K being small compared to the standard deviation in DIC among chains (Table S3). Consensus population assignments for K = 8, 9 and 10 were obtained for the five chains with the highest likelihood using CLUMPP (V1.1.2) [35] with parameters "-m 3 -w 0 -s 2" and with greedy option = 2 and repeats = 10,000. The similarity among the five chains (H') was 0.995 for K = 9. Compared to K = 9, populations 6 (African, S. E. Asia/Palm, Cocoa, Fruit) and 7 (Israel/Soil) were merged for K = 8, and a new population was inferred within populations 3 (Asian/Food, Drink) and 6 (African, S. E. Asia/Palm, Cacao, Fruit) for K = 10 (Figure S1). Multidimensional scaling was performed on all 5868 sites and 262 strains using the identity by state distance between each pair of strains and the "cmdscale" function in R with three dimensions. Hierarchical clustering of either sites or strains was performed using the "hclust" function in R with complete linkage and the euclidean distance of identity by state.

## Acknowledgements

We thank Meridith Blackwell, Andreas Hellström, Eviatar Nevo, Mat Goddard and Lene Jesperson for providing strains. We thank Eric Jeffery for help with the manuscript, Adrian Scott for help with the figures, and Scott Bloom for assistance with Illumina sequencing. A.M.D. is funded by a strategic partnership between ISB and the University of Luxembourg. J.C.F. is supported by National Institutes of Health grant GM080669.

## Figure Legends

**Figure 1.** Neighbor-joining tree of the 262 *S. cerevisiae* strains based on multiple alignment of 116,880 bases. Branch lengths are proportional to sequence divergence measured as P-distance. Scale bar indicates 5 polymorphisms/ 10kb of sequence. Geographical and environmental clusters of strains are named and are indicated by black-outlined/grey-filled ovals. Colored ovals with numbering refer to strain populations identified in Figure 2. Seven strains widely used in the laboratory are labeled.

**Figure 2.** Clustered genotypes with inferred population structure and membership. Sites were clustered by complete heirarchical clustering using the euclidean distance of allele sharing (identity by state). Strains were grouped by population structure and memberships inferred using InStruct. Minor alleles are shown in red, heterozygous sites in yellow, common alleles in black, missing data is gray. Populations are labeled by the most common source and/or geographic location from which they were originally isolated.

**Figure 3.** Coincidence of admixture between pairs of populations. Each bar shows the number of strains with at least 20% ancestry from a reference population (bar labels) and 20% ancestry with another population (indicated by color in the legend). For comparison, grey filled circles show the number of strains with more than 80% ancestry from each population.

**Figure 4.** Relatedness among strains and the inferred populations to which they belong. The first and second principal coordinates (A) and the first and third principal coordinates (B) obtained from multidimensional scaling. Each circle shows a strain with color indicating the population contributing the largest proportion of ancestry and size indicating the proportion of ancestry from that population (see legend). Circles ringed in black show strains with more than 20 heterozygous sites.

**Figure 5.** Subpopulations defined by clustering of low frequency alleles. Two-dimensional hierarchical clustering of low frequency sites and strains. InStruct assignments are shown on the left, clustered genotypes are shown in the middle, with minor alleles in red, heterozygous sites in yellow, common alleles in black, and missing data in gray. Selected subpopulations are labeled on the right.

**Supporting Information**

**Table S1.** Strains used in this study, with population assignments inferred by InStruct.

**Table S2.** Populations inferred using InStruct and summary statistics.

**Table S3.** Fit of the population structure model as a function of the number of populations.

**Figure S1.** Population ancestry of strains inferred by InStruct. Populations are color-coded and the proportion of population ancestry assigned to each strain is indicated by bar height. Strain ancestry is shown assuming 8, 9 and 10 populations (K), with the order of strains based on K = 9 and color-coding of major populations matching that of K = 9.

**Supplemental Dataset 1.** Matrix of polymorphic sites. The matrix consists of 5,868 biallelic sites (columns) and 262 strains (rows) with column labels indicating the chromosome number and position separated by a period. Genotypes are represented by 0 or 2 for homozygotes, 1 for heterozygotes and -9 for missing data. Entries are comma delimited.

**Supplemental Dataset 2.** Neighbor-joining tree of 262 *S. cerevisiae* strains based on multiple alignment of 116,880 bases in Newick format. This tree is a version of Figure 1 that includes strain labels and the maximum group membership from Figure 2.

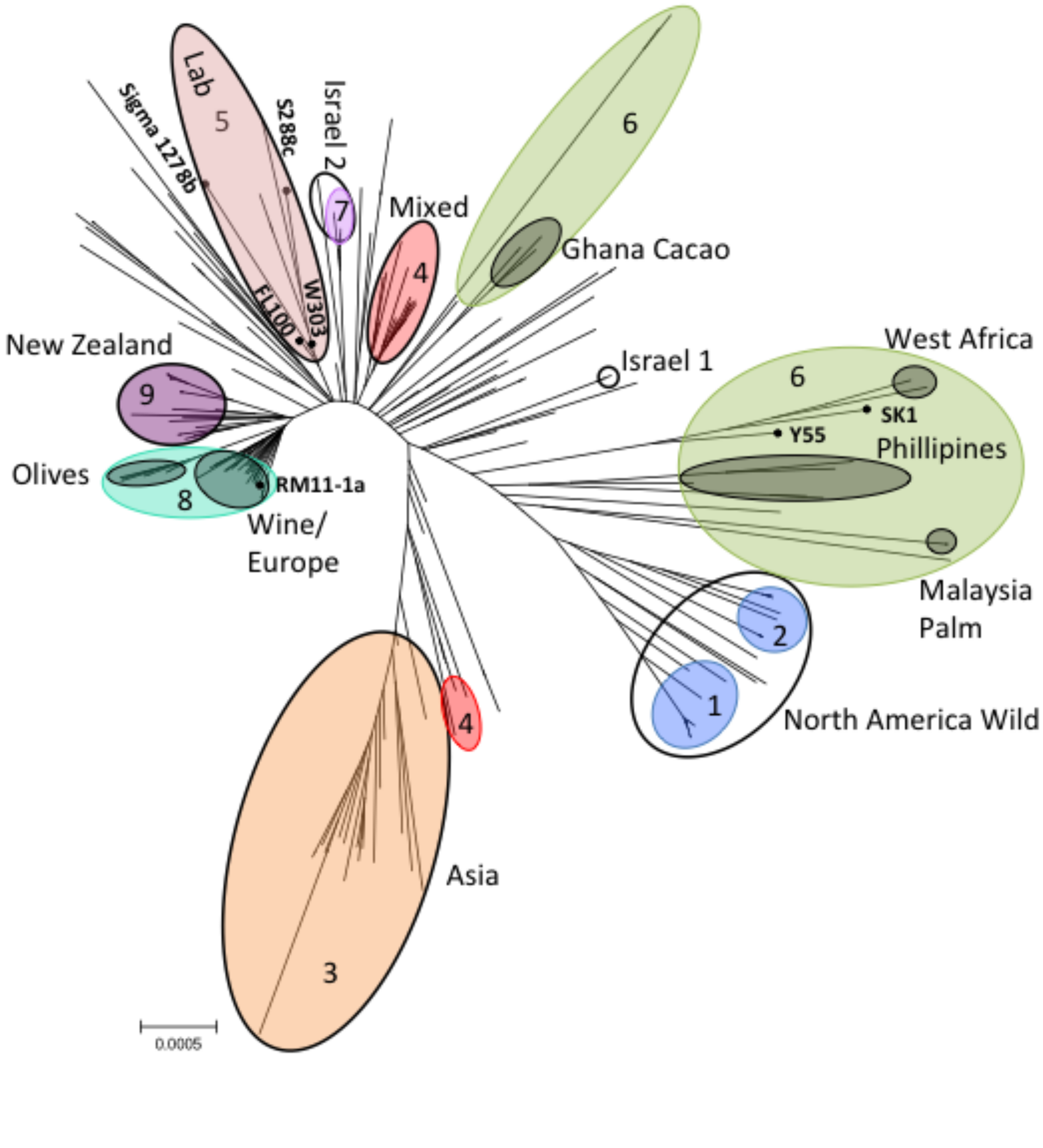

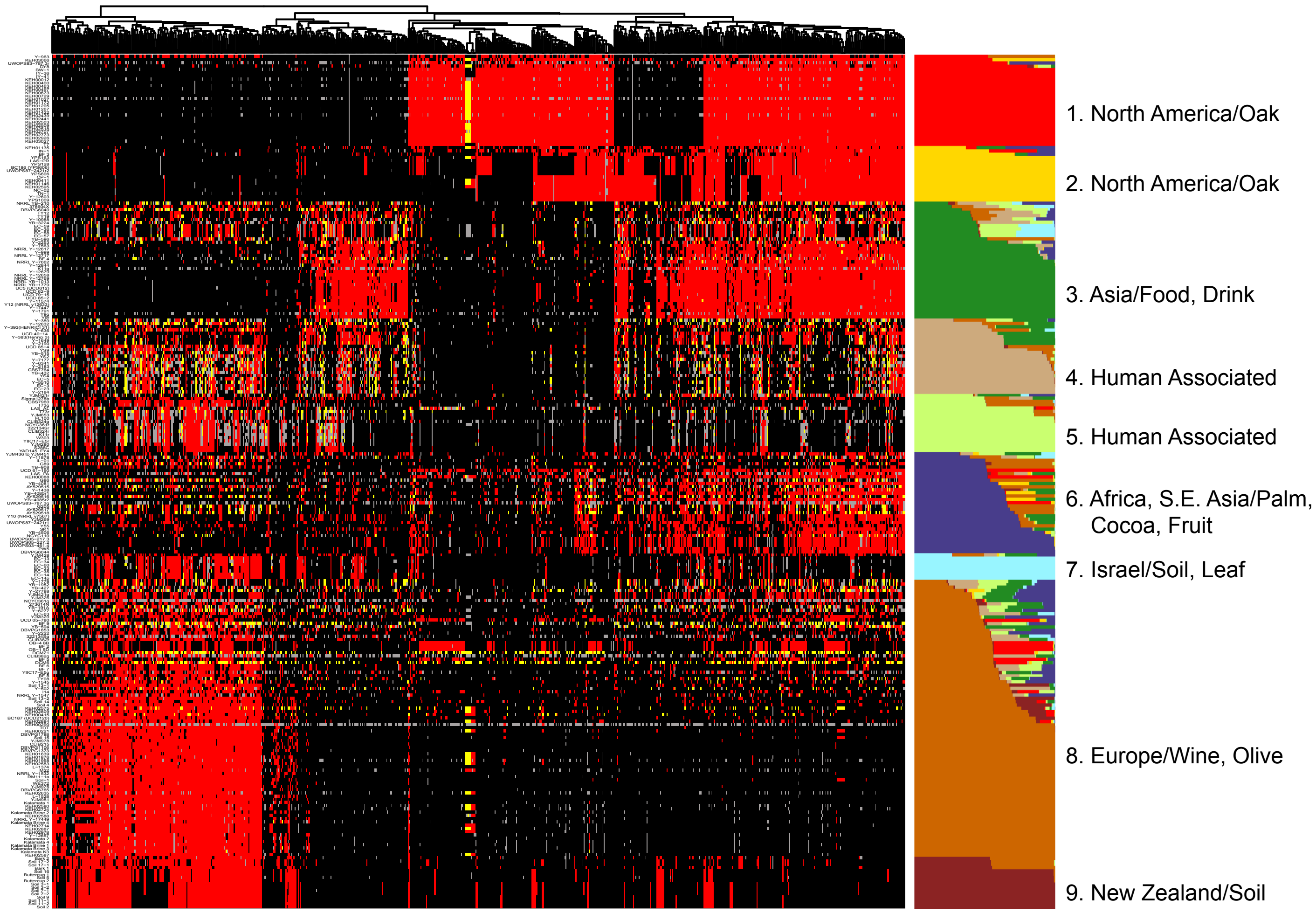

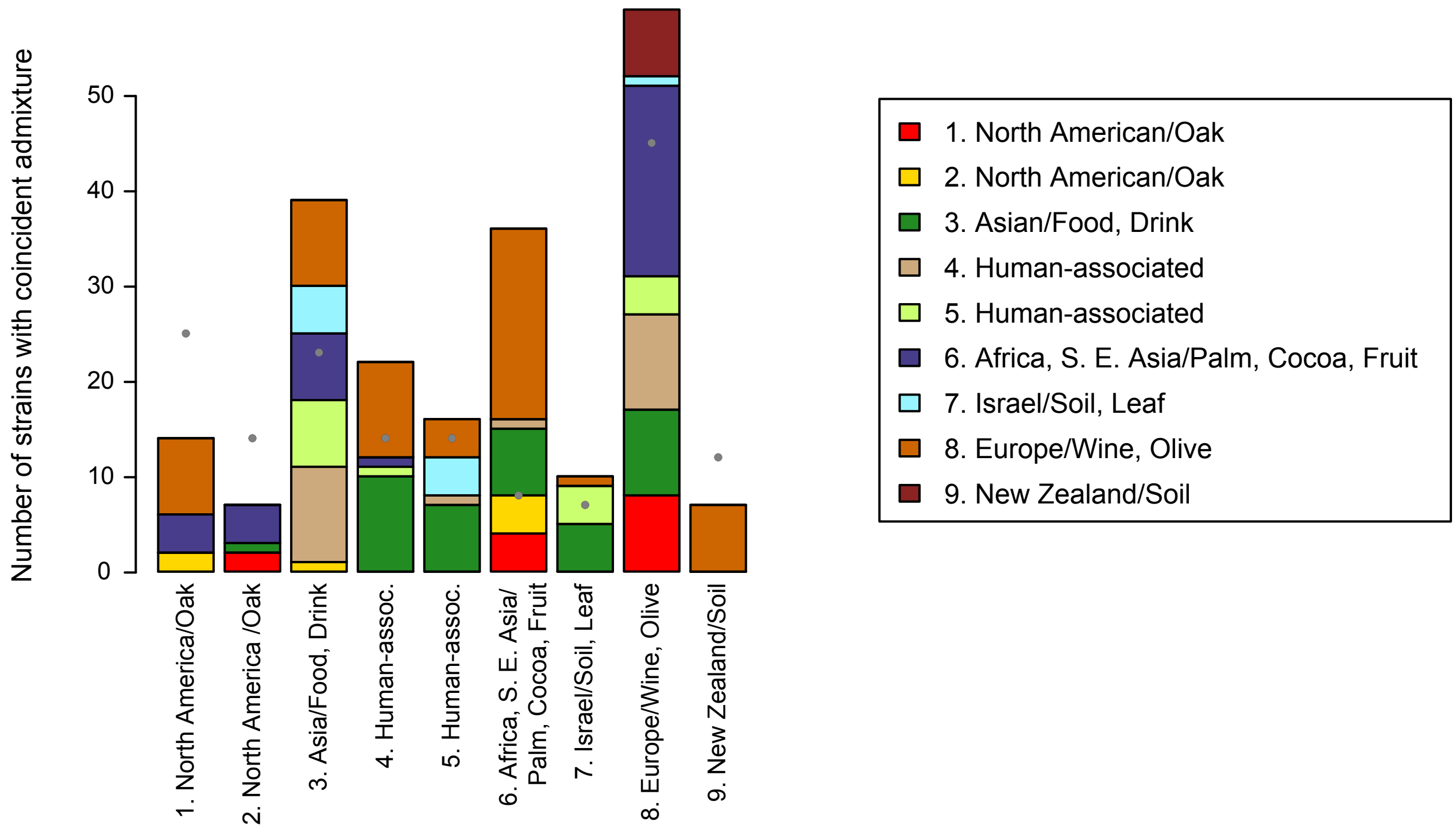

Number of strains with coincident admixture
0
10
20
30
40
50
1. North America/Oak
2. North America /Oak
3. Asia/Food, Drink
4. Human-assoc.
5. Human-assoc.
6. Africa, S. E. Asia/ Palm, Cocoa, Fruit
7. Israel/Soil, Leaf
8. Europe/Wine, Olive
9. New Zealand/Soil
1. North American/Oak
2. North American/Oak
3. Asian/Food, Drink
4. Human-associated
5. Human-associated
6. Africa, S. E. Asia/Palm, Cocoa, Fruit
7. Israel/Soil, Leaf
8. Europe/Wine, Olive
9. New Zealand/Soil

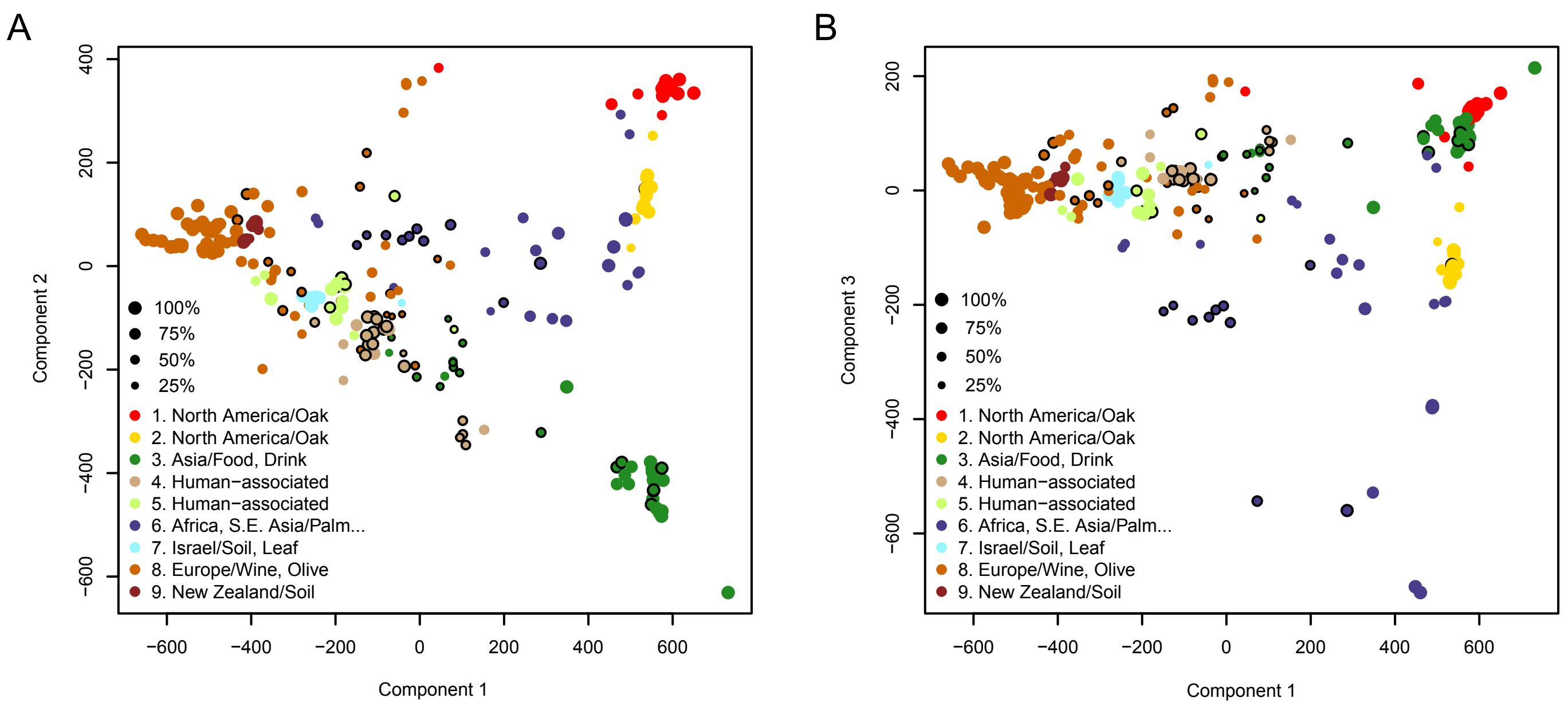

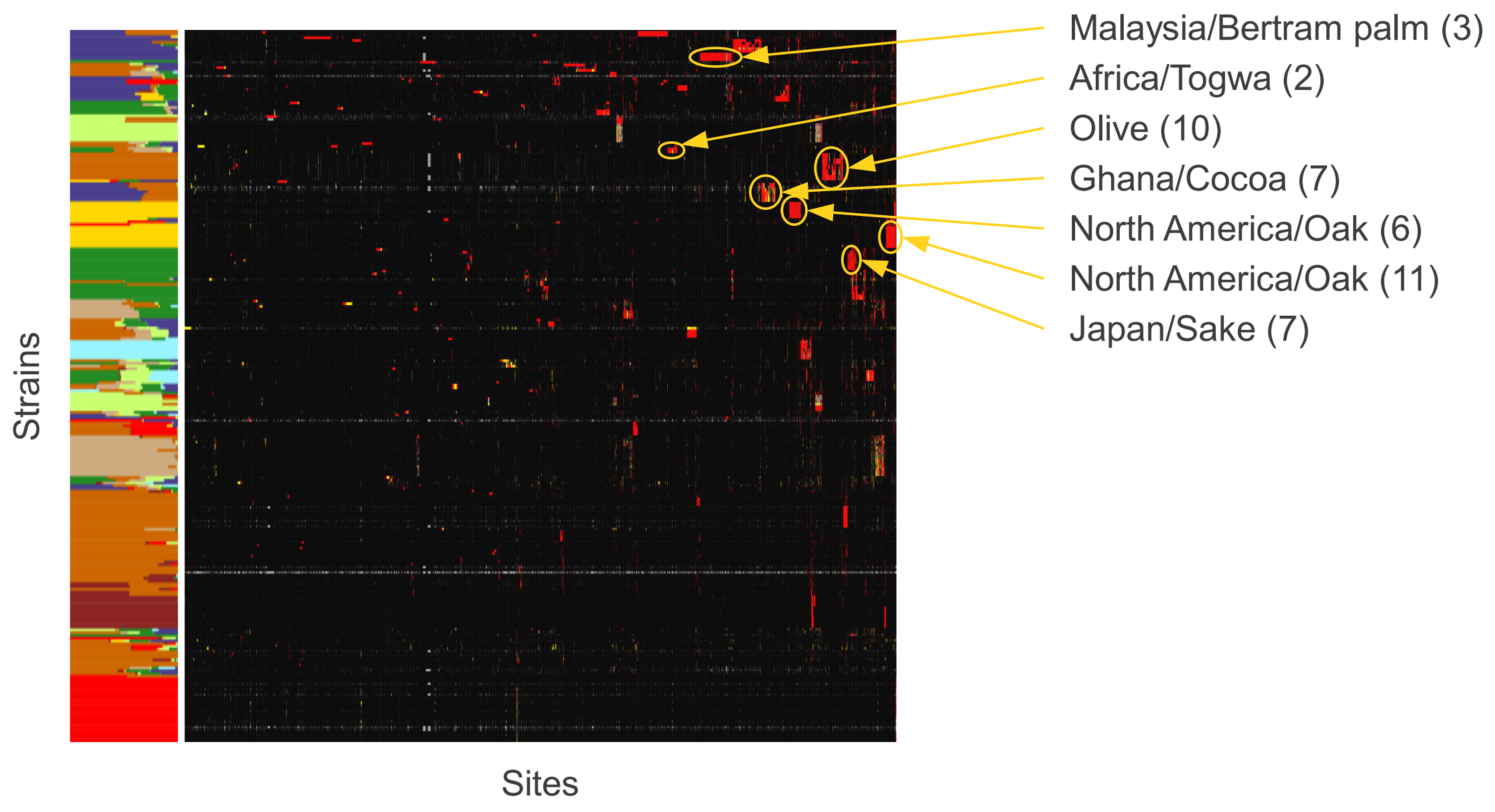

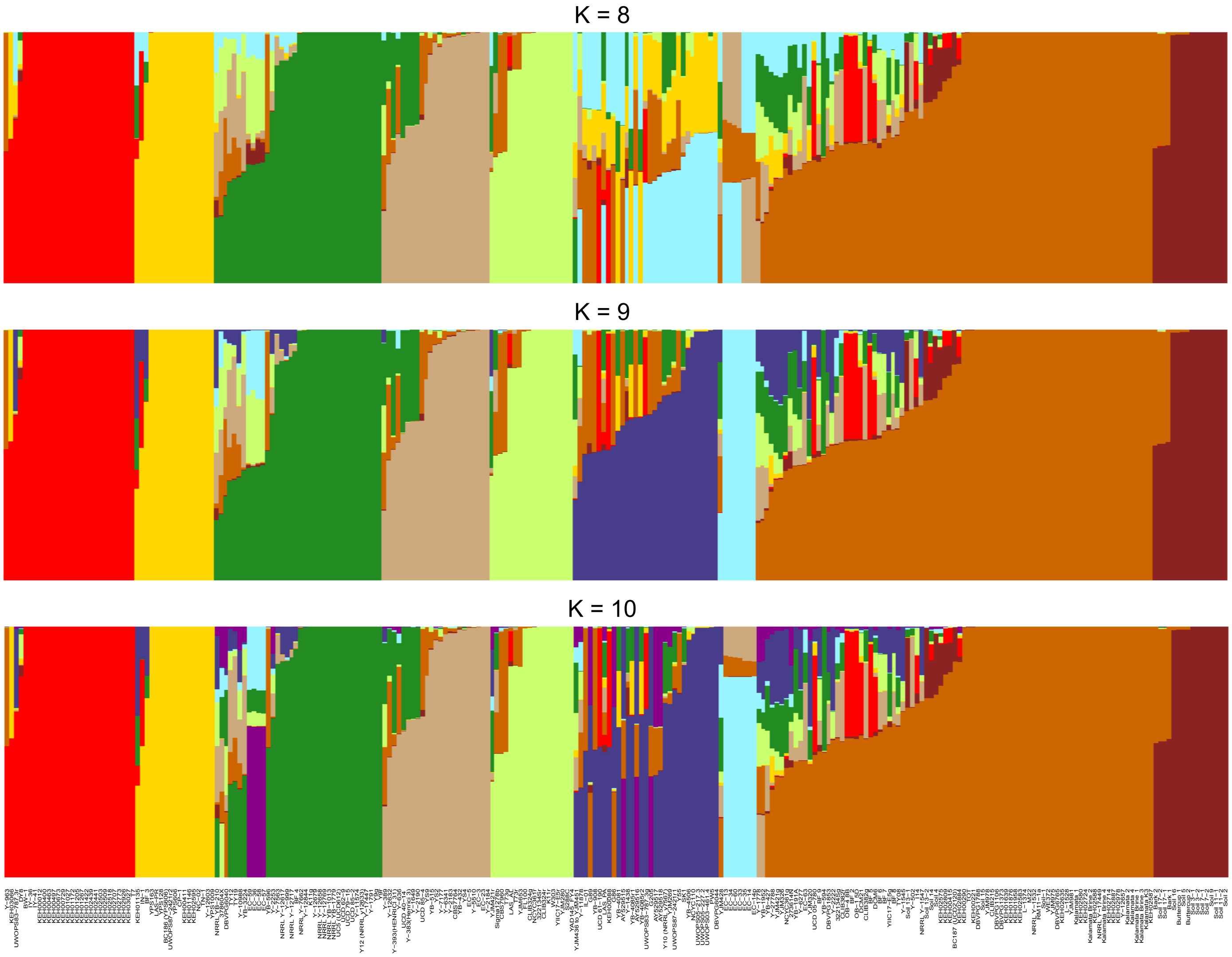
K = 8
K = 9
K = 10